\documentclass[12pt]{article}
\usepackage{graphicx}
\usepackage{amsmath,amssymb}

\newcommand\pubdate{\today}

\newcommand\pubnumber{change/delete REPORT-\#}

\def\Title#1{\begin{center} {\Large #1 } \end{center}}
\def\Author#1{\begin{center}{ \sc #1} \end{center}}
\def\Address#1{\begin{center}{ \it #1} \end{center}}

\newcommand\pubblock{\rightline{\begin{tabular}{l} \pubnumber\\
         \pubdate  \end{tabular}}}
\newenvironment{Abstract}{\begin{center}{\bf Abstract}\end{center} \bigskip \begin{quotation}  }{\end{quotation}}
\newenvironment{Presented}{\begin{quotation} \begin{center} 
             PRESENTED AT\end{center}\bigskip 
      \begin{center}\begin{large}}{\end{large}\end{center} \end{quotation}}

\def\beq{\begin{equation}}
\def\eeq#1{\label{#1}\end{equation}}
\def\eeqn{\end{equation}}

\def\beqa{\begin{eqnarray}}
\def\eeqa#1{\label{#1}\end{eqnarray}}
\def\eeqan{\end{eqnarray}}

\let\bar=\overbar

\def\Dslash{\not{\hbox{\kern-4pt $D$}}}
\def\dslash{\not{\hbox{\kern-2pt $\del$}}}

\def\BR{\mbox{\rm BR}}

\def\msb{{\bar{\ssstyle M \kern -1pt S}}}

\input definitions.tex

\begin{document}
\begin{titlepage}
\pubblock

\vfill


\Title{Search for \CP violation in \mtau and $D$ decays with a \KS in the final state.}
\vfill
\Author{M. Martinelli\footnote{now at {\it NIKHEF, National Institute for Nuclear Physics and High Energy Physics, NL-1009 DB Amsterdam, The Netherlands}.}\\On behalf of the \babar\ Collaboration.}
\Address{Universit\`a degli Studi di Bari and INFN, Bari, 70124, ITALY\\
SLAC National Accelerator Laboratory, Stanford, California 94309 USA}
\vfill


\begin{Abstract}
I report the recent searches for \CP violation in \mtau and $D$ decays including a \KS in the final state.
The analyses herein shown are based on data samples recorded by \babar\ and {\it\sc Belle} experiments.
A brief introduction on \CP violation is followed by the summary of the experimental techniques and the results
obtained for \mtau and $D$ decays, respectively.
Finally, an outlook on future development is provided.
\end{Abstract}

\vfill

\begin{Presented}
The Ninth International Conference on\\
Flavor Physics and CP Violation\\
(FPCP 2011)\\
Maale Hachamisha, Israel,  May 23--27, 2011
\end{Presented}
\vfill

\end{titlepage}
\def\thefootnote{\fnsymbol{footnote}}
\setcounter{footnote}{0}
%


\section{Introduction}

The violation of the charge-conjugation and parity simmetry (\CP) is introduced in the Standard Model (SM)
by the Kobayashi-Maskawa phase in the Cabibbo-Kobayashi-Maskawa quark mixing matrix~\cite{CKM}.
This effect usually results into an asymmetry among the decay rates of a process and its charged-conjugate.
\CP violation is one of the most intriguing features of the SM, and has lead to many advancement in particle physics
since it has been discovered in kaon decays~\cite{Christenson:1964fg}.
The effect of \CP violation has been extensively studied in kaon and \B decays~\cite{CP:KBres,Abe:2001xe,Bigi:2000yz}, with results 
that deeply influenced our knowledge of physics~\cite{Sakharov:1967dj}.

Recently, there is a vivid interest on the search for \CP violation in different frameworks, such as charm meson and lepton decays, due to the higher sensitivities reached by the \B factories.
In these cases, the \CP violating effect is predicted to be extremely small or absent in the SM.
Observing a signal would then strongly indicate the presence of processes generated by physics beyond the SM.

\section{\CP violation in \mtau decays}

In the SM, no \CP violation is expected in lepton decays, except for those including a \Kz, due to the \KL-\KS mixing 
and the \CP violating phase introduced by the \KL\to\pipi decay.
The \CP violation is then introduced by the detection of kaons having short decay times.
The expected amount of \CP violation is 0.3\%, obtained from the measurements made on \KL\to\pipi~\cite{BigiSandaTau}.

Aside the SM \CP violation, there is a New Physics (NP) model that provides \CP violation in \mtau decays.
In the Multi-Higgs-Doublet-Model, the addition of a charged scalar Higgs boson can introduce \CP violation in the 
angular distribution of \mtau decays.
The effect of the introduction of a new boson can be described modifying the scalar form factor of the hadronic functions 
describing the decay process
\begin{align}
F_S(Q^2) \to \tilde{F}_S(Q^2)= F_S(Q^2) + \frac{\eta_S}{m_{\tau}}F_H(Q^2),
\end{align}
where $\eta_S$ is an adimensional complex coupling costant.
Referring to~\cite{KuhnMirkes,Weinberg,Grossman}, the \CP violating observable is related to $\eta_S$
\begin{align}
A^{\CP} = \frac{\int\cos\beta\cos\psi\left(\frac{d\Gamma_{\taum}}{d\omega} - \frac{d\Gamma_{\taup}}{d\omega}\right)d\omega}{\frac{1}{2}\int{\left(\frac{d\Gamma_{\taum}}{dQ^2}+\frac{d\Gamma_{\taup}}{dQ^2}\right)dQ^2}}\simeq c_i\Im(\eta_S)\qquad \left(d\omega=dQ^2d\cos\beta d\cos\phi\right),
\end{align}
where $\beta$ and $\psi$ are the polar angles, $d\Gamma_{\tau}$ is the decay rate and $Q^2$ is the energy.

\subsection{Experimental results}
\begin{figure}[htb]
\centering
\includegraphics[height=0.4\textwidth]{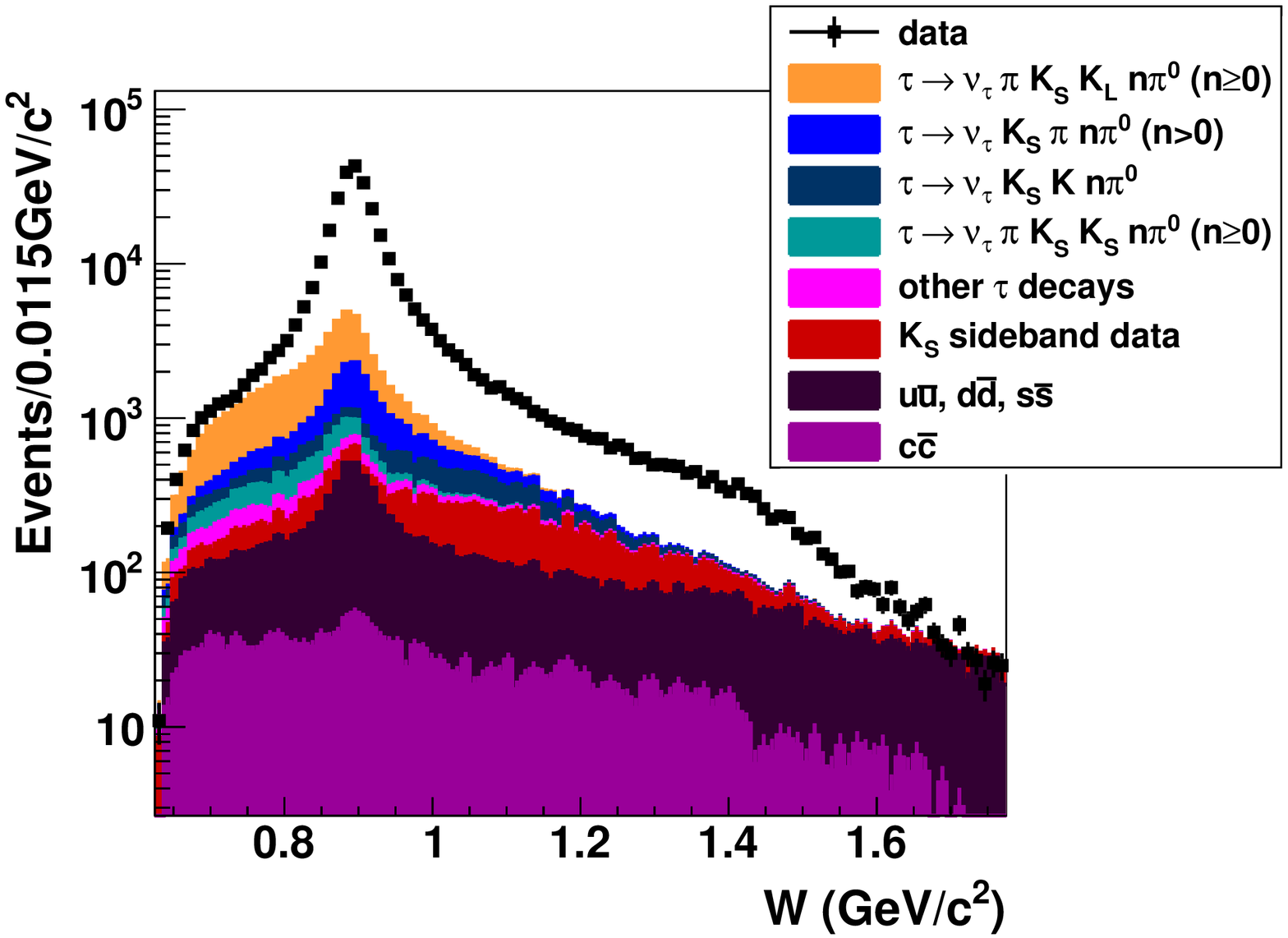}
\includegraphics[height=0.4\textwidth]{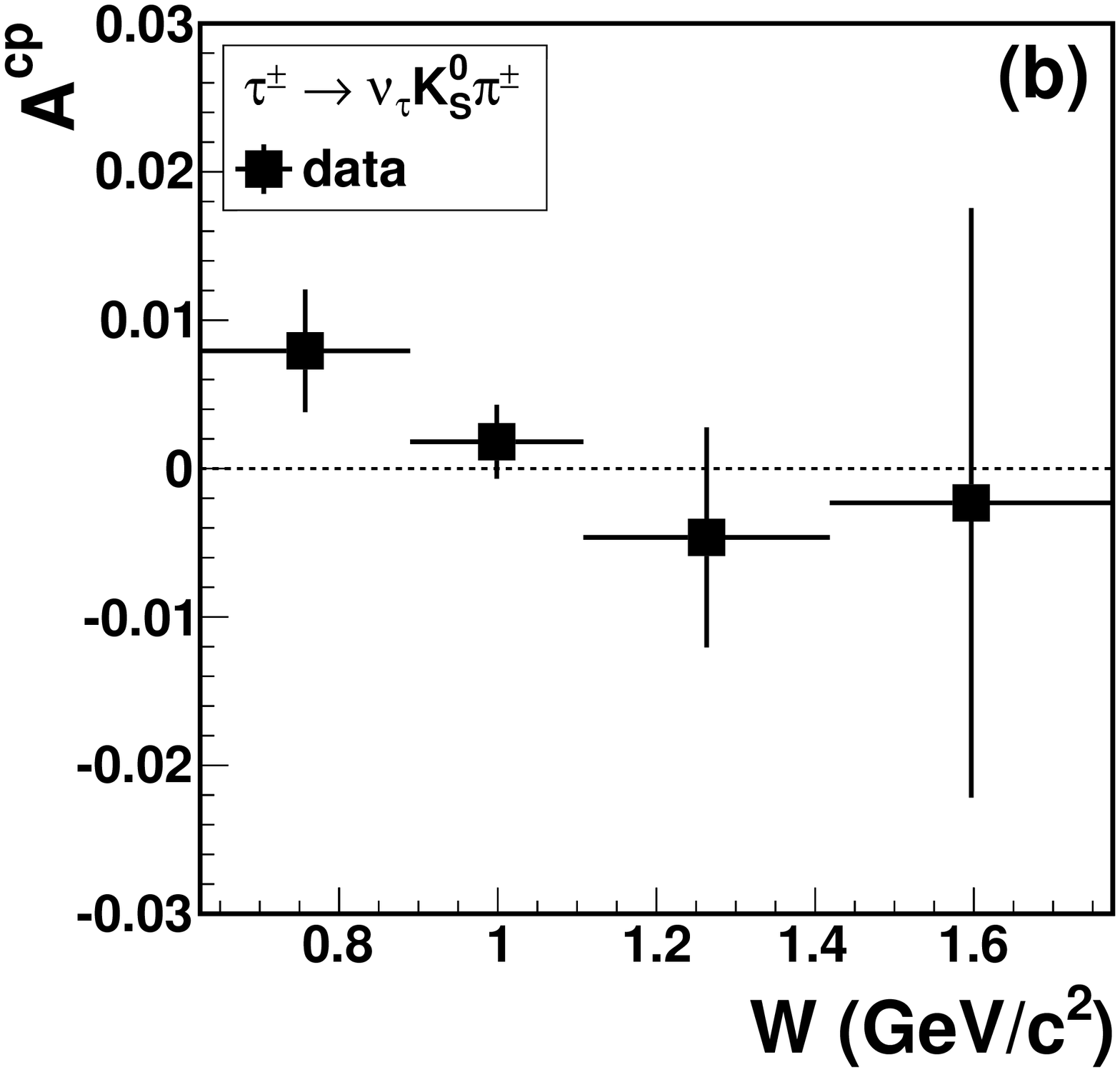}
\caption{The background contributions estimated from Monte Carlo (left) and the measured asymmetry parameters in 
bins of $W$ (right).}
\label{fig:tau}
\end{figure}

The {\sc Belle} Collaboration recently submitted a search for \CP violation from NP using \mtau\to\KS\mpi\nut 
decays~\cite{Bischofberger:2011pw}.
They measured $A^{\CP}$ in four bins of $W=m(\KS\mpi)$
\begin{align}
A^{\CP} \simeq \frac{\left<\cos\beta\cos\psi\right>_{\taum}}{1-f_b^-} - \frac{\left<\cos\beta\cos\psi\right>_{\taup}}{1-f_b^+}, 
\end{align}
where $f_b^{\pm}$ is the fraction of background evaluated from studies on Monte Carlo.
The background contributions are shown in Fig.~\ref{fig:tau}. 
In the same figure, the asymmetry parameter measured in bins of $W$ is shown. 

As can be observed from the figure, $A^{\CP}$ is consistent to zero for each bin, except for the lower mass one, where a $1.9\sigma$ deviation is found.
From this measurement a limit on the amplitude of $\Im(\eta_S)$ can be set: $|\Im(\eta_S)|<(0.012-0.026)@90\%~C.L.$, 
that improves of an order of magnitude the previous result from CLEO~\cite{Bonvicini:2001xz}.

\section{\CP violation in $D$ decays}
The \babar\ and {\sc Belle} experiments are now allowed to probe \CP violation in charm decays with sensitivity of the order of $10^{-3}$, the same order of magnitude of the higher SM expectations.
In the mean time, NP can contribute up to the percent level, making it clearly distinguishable from SM.
Among the others, the NP models that introduce \CP violation through one-loop processes, such as QCD penguins 
and dipole operators, or flavor changing neutral currents in supersymmetric flavor models, are the most probable to show
the large effect, if present~\cite{Grossman:2006jg,Buccella:1994nf}.

Similarly to what observed before, the presence of the \KS in the final state introduces a further \CP violation asymmetry
of about 0.3\% due to interference between \Kz-\Kzb mixing and \KL \CP violation.
This effect needs to be taken into account when discussing the final result. 

\subsection{Direct \CP violation measurements}
The more intuitive way to measure \CP violation is to directly compare the $D$ and \Db decay rates
\begin{align}
A_{\CP}^{\text{rec}}=\frac{\Gamma_D - \Gamma_{\Db}}{\Gamma_D+ \Gamma_{\Db}}.
\label{eq:acpRec}
\end{align}
However, this observable does not provide a valid measurement of \CP violation.
In asymmetric detectors, like \babar\ and {\sc Belle}, one indeed needs to consider the effect of forward-backward
asymmetry ($A_{FB}$) generated by the interference between the electro-weak and the electro-magnetic \epem\to\ccbar
production processes~\cite{Halzen:1984mc}.
This effect is a function of the cosine of the \ccbar production polar angle.

Another trivial source of asymmetry is generated by the different interaction between particles and the detector 
($A_{\epsilon}$).
Oppositely charged particles have indeed different cross-section when reacting to the same detector material.
If the final state does not have the same number of particles and anti-particles, this effect can influence the asymmetry.

In the end, the general expression to describe the \CP violating asymmetry observable shown in Eq.~\ref{eq:acpRec} is
\begin{align}
A_{\CP}^{\text{rec}} = A_{\CP}+A_{FB}+A_{\epsilon}.
\end{align}
One than needs to isolate $A_{FB}$ and $A_{\epsilon}$ in order to measure the \CP violating observable $A_{\CP}$.

The first search of this kind in \Dps decays involving a \KS has been made by {\sc Belle}~\cite{Ko:2010ng}, that analyzed \Dps\to\KS\pip and \Dps\to\KS\Kp decays.
When studying \Dps\to\KS\pip decays, the contribution of $A_{FB}$ and $A_{\epsilon}^{\pip}$ is evaluated from a 
control sample of \Ds\to\mphi\pip.
The control sample has been subtracted bin per bin into the phase space of $p^{\text{lab}}_{\pi}$, 
$\cos\theta^{\text{lab}}_{\pi}$ and $\cos\theta^{\text{CMS}}_D$, that are the momentum and the cosine of the polar
angle of the pion in the laboratory frame and the cosine of the polar angle of the $D$ meson in the center-of-mass frame, 
respectively.
After weighting (integrating for \Ds) on the phase space, the \CP violation parameter for \Dps\to\KS\pip is 
\begin{align}
A_{\CP}^{\Dp\to\KS\pip} &= (-0.71\pm 0.26\pm 0.20)\%,\\
A_{\CP}^{\Ds\to\KS\pip} &= (+5.45\pm 2.50\pm 0.33)\%,
\end{align}
consistent with the SM expectations of -0.3\%.

In the \Dps\to\KS\Kp analysis, one needs to correct for the \Kp reconstruction asymmetry, but there are no control samples
that allow to evaluate $A_{FB}$ and $A_{\epsilon}^{\Kp}$ at the same time.
The solution is to evaluate first $A_{\epsilon}^{\Kp}$ as the difference between $A_{\text{rec}}^{\Dz\to\Km\pip}$ and
$A_{\text{rec}}^{\Ds\to\mphi\pip}$, then the two remaining contributions to $A_{\CP}^{\text{rec}}$, $A_{\CP}$ and $A_{FB}$,
are separated by considering that the former is even respect to $\cos\theta^*_D$, while the latter is odd
\begin{align}
A_{\CP}(\cos\theta^*_D) &= \frac{A_{\text{rec}}^{\text{corr}}(|\cos\theta^*_D|) + A_{\text{rec}}^{\text{corr}}(-|\cos\theta^*_D|)}{2},\label{eq:ACPcos}\\
A_{FB}(\cos\theta^*_D) &= \frac{A_{\text{rec}}^{\text{corr}}(|\cos\theta^*_D|) - A_{\text{rec}}^{\text{corr}}(-|\cos\theta^*_D|)}{2}.
\label{eq:AFBcos}
\end{align}

A weighted average on five bins of $|\cos\theta^*_D|$ gives
\begin{align}
A_{\CP}^{\Dp\to\KS\Kp} &= (-0.16\pm 0.58\pm 0.25)\%,\\
A_{\CP}^{\Ds\to\KS\Kp} &= (+0.12\pm 0.36\pm 0.22)\%,
\end{align}
consistent to the SM expectations.

The \babar\ Collaboration slightly improved the technique in the analysis of \Dp\to\KS\pip 
decays~\cite{delAmoSanchez:2011zz}.
The $A_{\epsilon}^{\pi}$ is here measured using a control sample of \BB decays and mapping the pion
reconstruction efficiency in bins of pion momentum and its polar angle in the laboratory frame.
This procedure has found to produce a bias of +0.05\% to $A_{\CP}$ that has been included in the systematics.
The improvement given by this technique is that the systematic error is limited to 0.1\%, that is actually the best result 
for a \CP violation analysis of $D$ meson decays.

\begin{figure}[htb]
\centering
\includegraphics[height=0.4\textwidth]{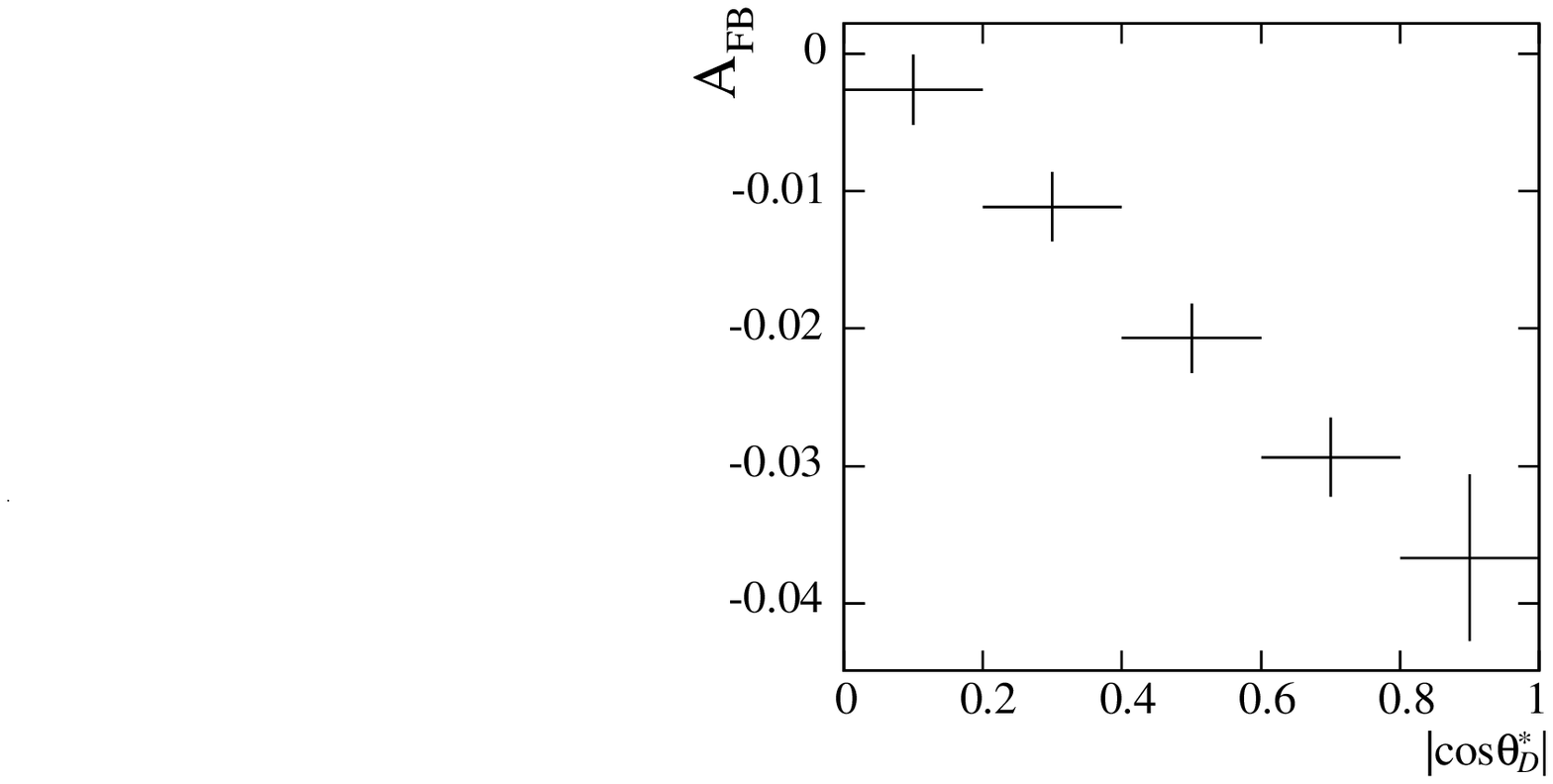}
\includegraphics[height=0.4\textwidth]{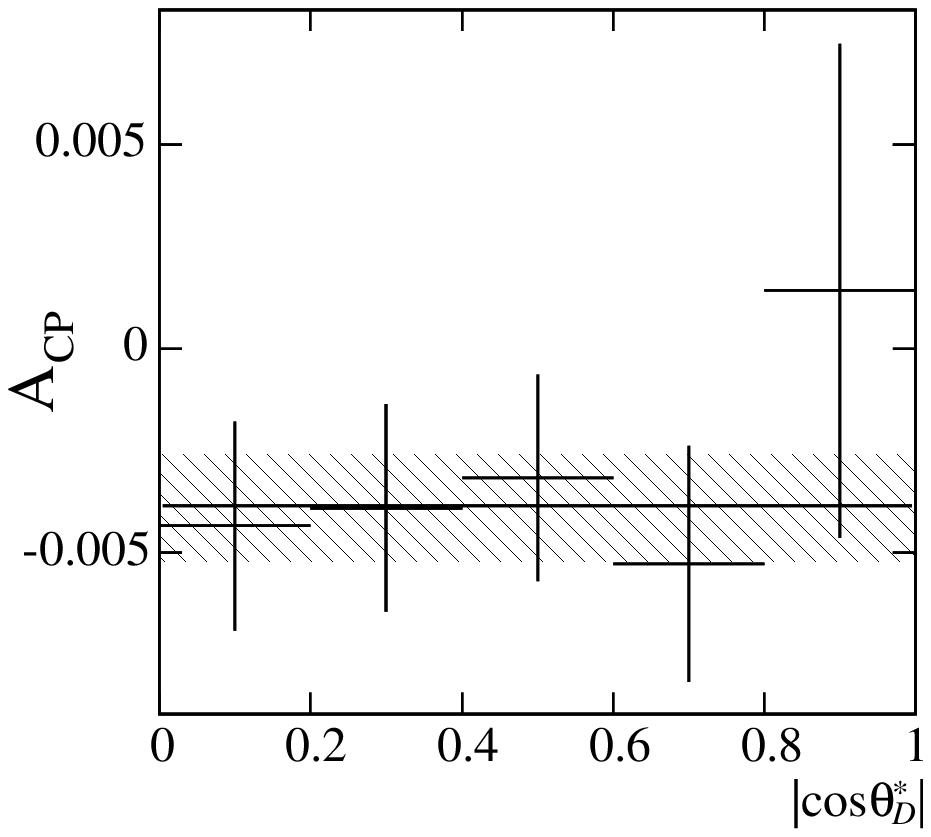}
\caption{\label{fig:DKSpiBabar}The forward-backward asymmetry $A_{FB}$ (left) and the \CP violating asymmetry $A_{\CP}$ (right) in bins of $|\cos\theta^*_D|$. 
The shaded area in the left end plot represents the averaged value of $A_{\CP}$ including a $\pm1\sigma$ error.}
\end{figure}

The asymmetry parameters, $A_{\CP}$ and $A_{FB}$ are measured using Eq.~\ref{eq:ACPcos} and~\ref{eq:AFBcos},
with the results shown in Fig.~\ref{fig:DKSpiBabar}.

The measurement of the \CP violating asymmetry is made averaging on the five bins of $|\cos\theta^*_D|$ and gives
\begin{align}
A_{\CP}^{\Dp\to\KS\pip} &= (-0.44\pm 0.13\pm 0.10)\%,
\end{align}
consistent to 0 at $2.7\sigma$ and to the SM prediction of -0.3\%.

Recently, also {\sc Belle} Collaboration managed to reduce the systematic error of this kind of analysis to the 0.1\% level.
They measured the \CP violation parameter in \Dz\to\KS\Pz (\Pz=\mpi,\meta,\etapr). 
In the same analysis, an indirect measurement of SM \CP violation from mixing and interference is provided.
The \CP violation parameter is evaluated from the asymmetry between \Dstarp and \Dstarm decays they reconstruct from \Dz\pip.
In order to isolate the \CP violation parameter, they evaluate the asymmetry introduced by the soft pion reconstruction ($A_{\pi_s}^{\text{rec}}$) using a control sample of tagged and untagged \Dz\to\Km\pip events and evaluate $A_{\pi_s}^{\text{rec}}=A_{\text{tagged}}^{\text{rec}}-A_{\text{untagged}}^{\text{rec}}$.
They separate \CP violation from forward-backward asymmetry using Eq.~\ref{eq:AFBcos} and obtain the following results:
\begin{align}
A_{\CP}^{\Dz\to\KS\mpi} &= (-0.28\pm 0.19\pm 0.10)\%,\\
A_{\CP}^{\Dz\to\KS\meta} &= (+0.54\pm 0.51\pm 0.16)\%,\\
A_{\CP}^{\Dz\to\KS\etapr} &= (+0.98\pm 0.67\pm 0.14)\%.
\end{align}

Finally, they use the result on $A_{\CP}^{\Dz\to\KS\mpi}$ to test for indirect \CP asymmetry universality:
\begin{align}
a^{ind}=A_{\CP}^{\Dz\to\KS\mpi}-A_{\CP}^{\KS}=(+0.05\pm0.19\pm0.10)\%,
\end{align}
consistent to $-A_{\Gamma}= (-0.14\pm0.27)\%$, obtained from \Dz\to\Kp\Km and \Dz\to\pip\pim measurements~\cite{Nakamura:2010zzi}.

\subsection{Measuring \CP violation using $T$-odd correlations}
\CP violation can be measured indirectly using $T$-odd correlations.
The presence of a $T$-odd correlation indicates $T$ violation and, assuming the validity of \CPT invariance, \CP violation can be inferred.
An observable for $T$-odd correlation can be easily built in four-body decays using the momentum of the particles in their mother's rest frame~\cite{Bigi:2001sg}.
In the case of $D$ decays, the suitable final states for such an analysis are \Dz\to\Kp\Km\pip\pim~\cite{delAmoSanchez:2010xj} and \Dps\to\Kp\KS\pip\pim~\cite{delAmoSanchez:2011aa}.

In both the two cases, the $T$-odd correlation observable can be written as
\begin{align}
C_T= \vec{p}_{\Kp}\cdot(\vec{p}_{\pip}\times\vec{p}_{\pim}),
\end{align}
and the $T$-odd correlation is found when the asymmetry
\begin{align}
A_T= \frac{\Gamma_{D}(C_T>0)-\Gamma_{D}(C_T<0)}{\Gamma_{D}(C_T>0)+\Gamma_{D}(C_T<0)}\neq0.
\end{align}
However this is not yet a signal for $T$, since Final State Interaction (FSI) may produce the asymmetry~\cite{Bigi:2009zzb}.
A simple solution is to measure the asymmetry parameter on the charged-conjugate decay and measure
\begin{align}
\mathcal{A}_T= \frac{1}{2}\left(A_T - \bar{A}_T\right),
\label{eq:Atv}
\end{align}
that is an asymmetry that characterizes $T$ violation in the weak decay process.

Recently, \babar\ submitted the measurements of this asymmetry in \Dp and \Ds decays to the final state \Kp\KS\pip\pim~\cite{delAmoSanchez:2011aa}.
The analysis considered inclusive $D$ decays, selected by means of particle identification, kinematic vertex fit and a likelihood ratio. The four datasets obtained by separating the sample depending on $D$ charge and $C_T(\bar{C}_T)$ value are fitted simultaneously to measure the asymmetry parameters for \Dp and \Ds, respectively. 
\begin{figure}[htb]
\centering
\includegraphics[width=0.22\textwidth]{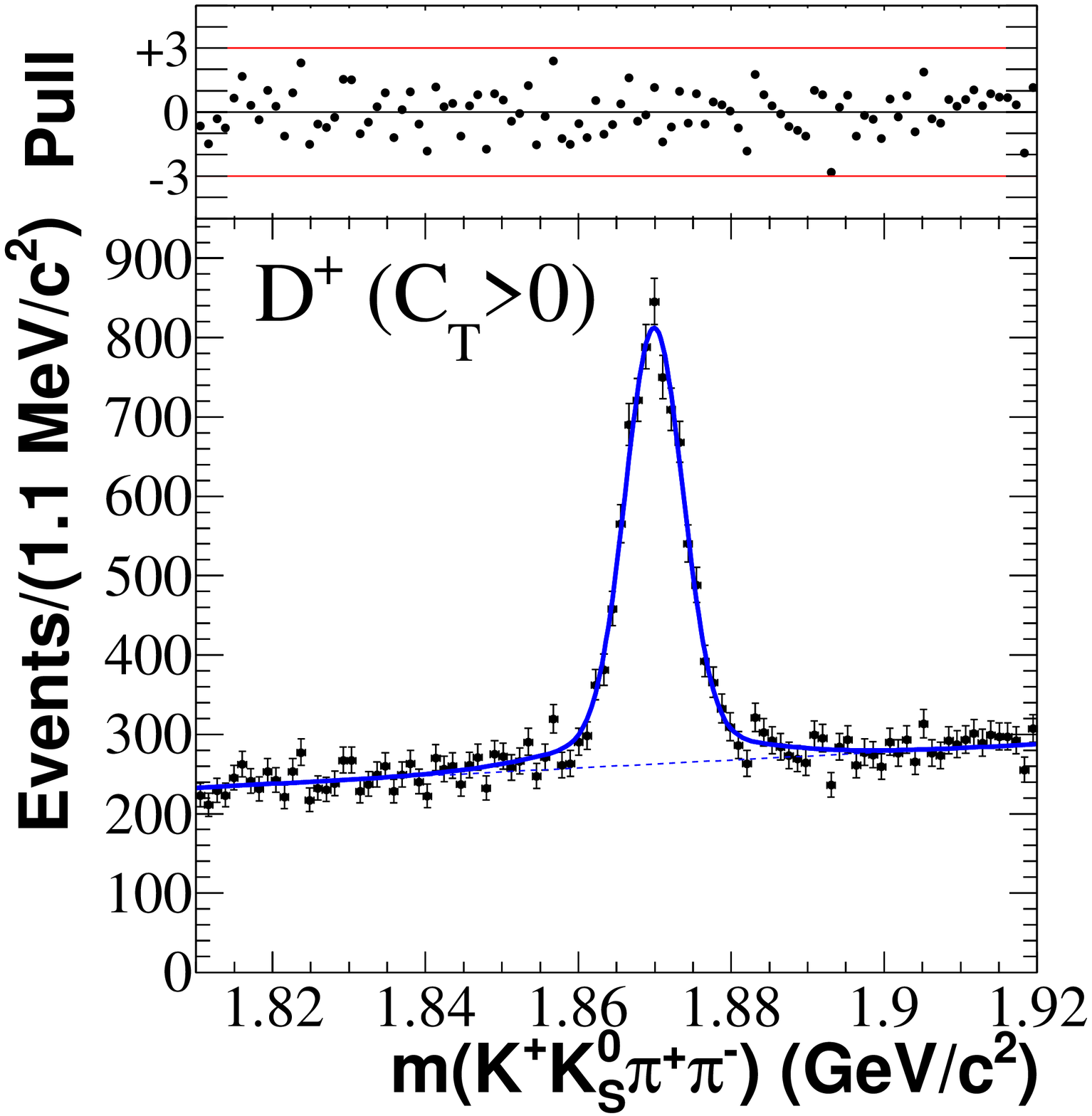}
\includegraphics[width=0.22\textwidth]{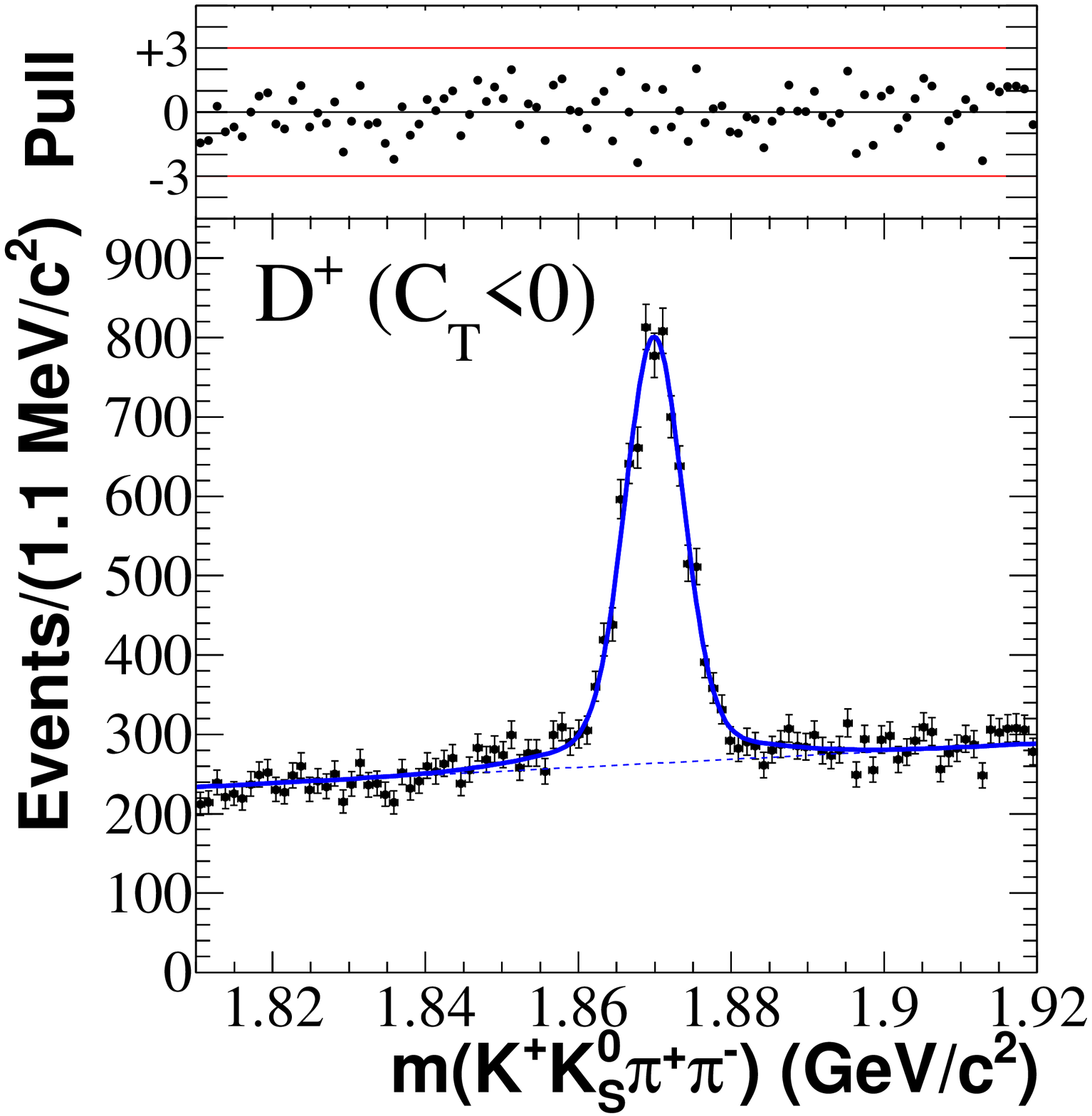}
\includegraphics[width=0.22\textwidth]{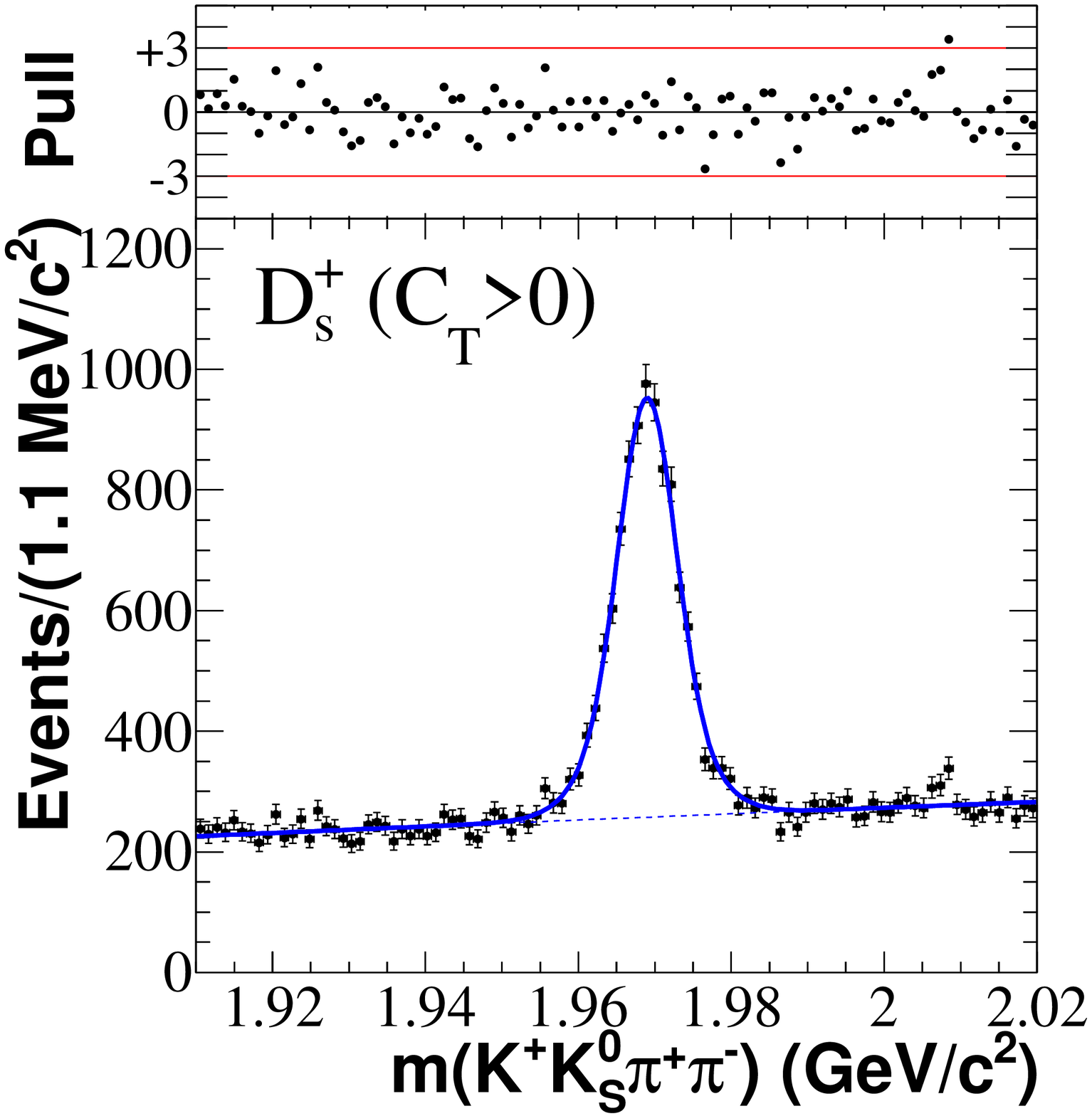}
\includegraphics[width=0.22\textwidth]{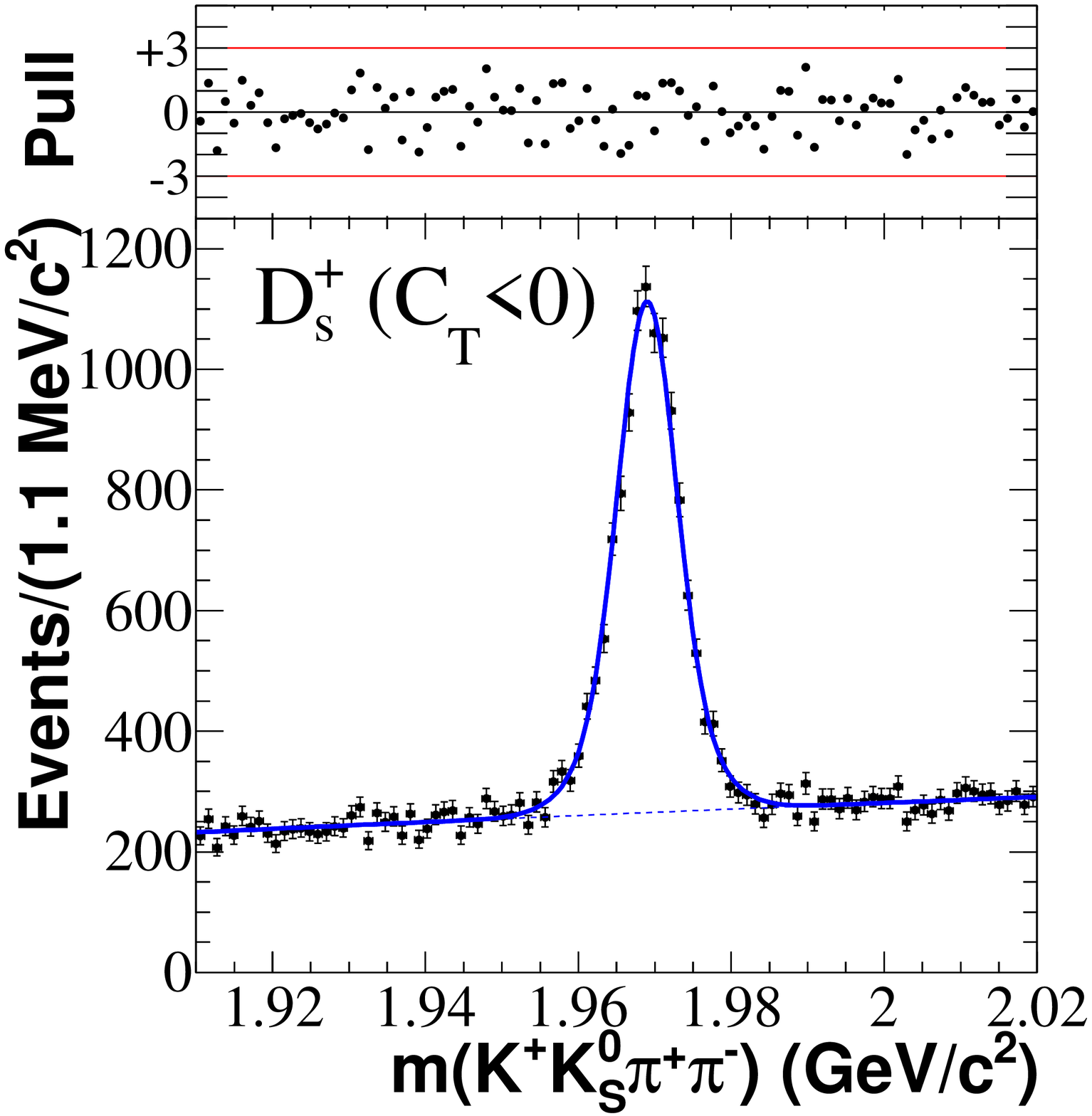}
\includegraphics[width=0.22\textwidth]{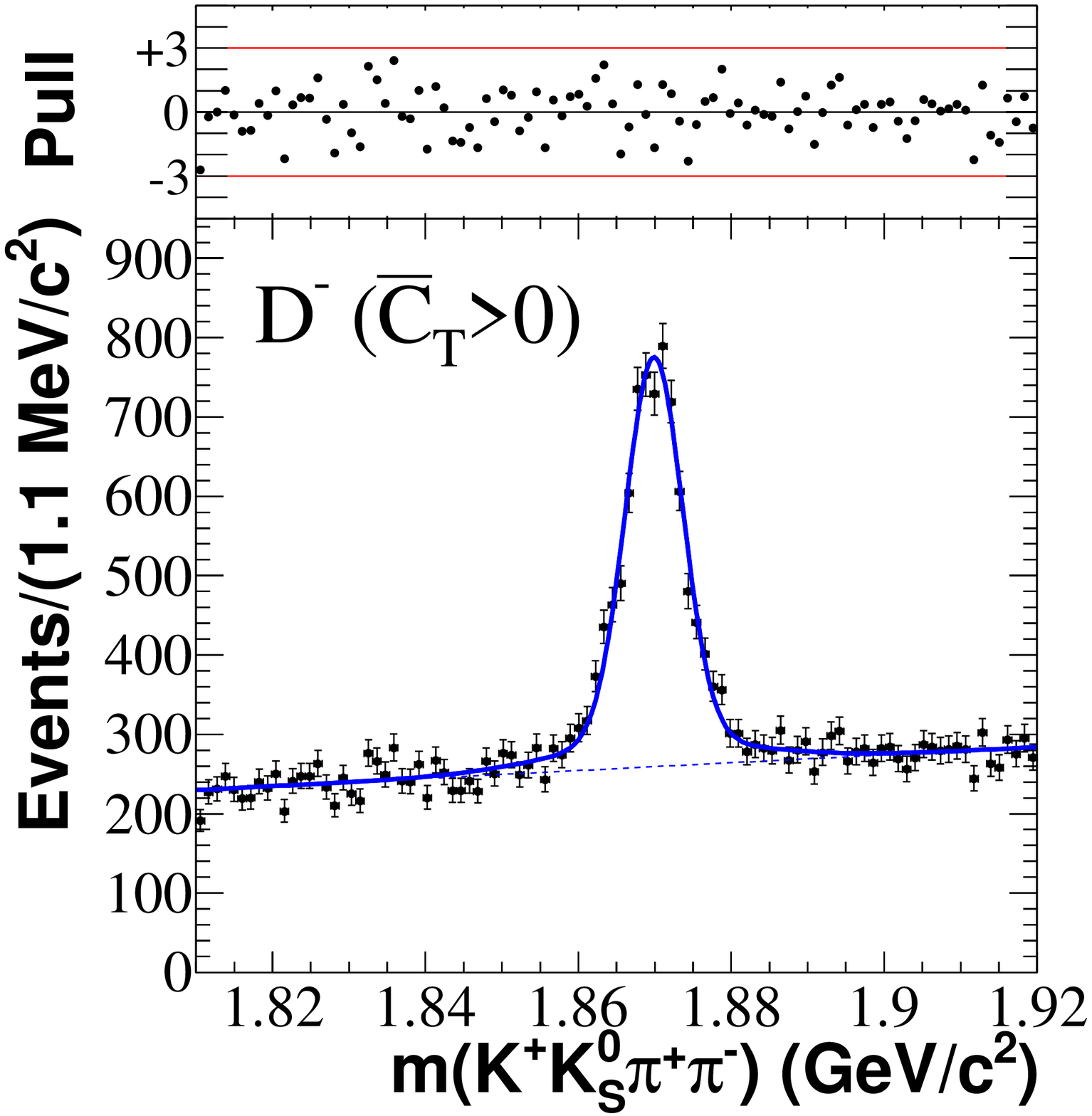}
\includegraphics[width=0.22\textwidth]{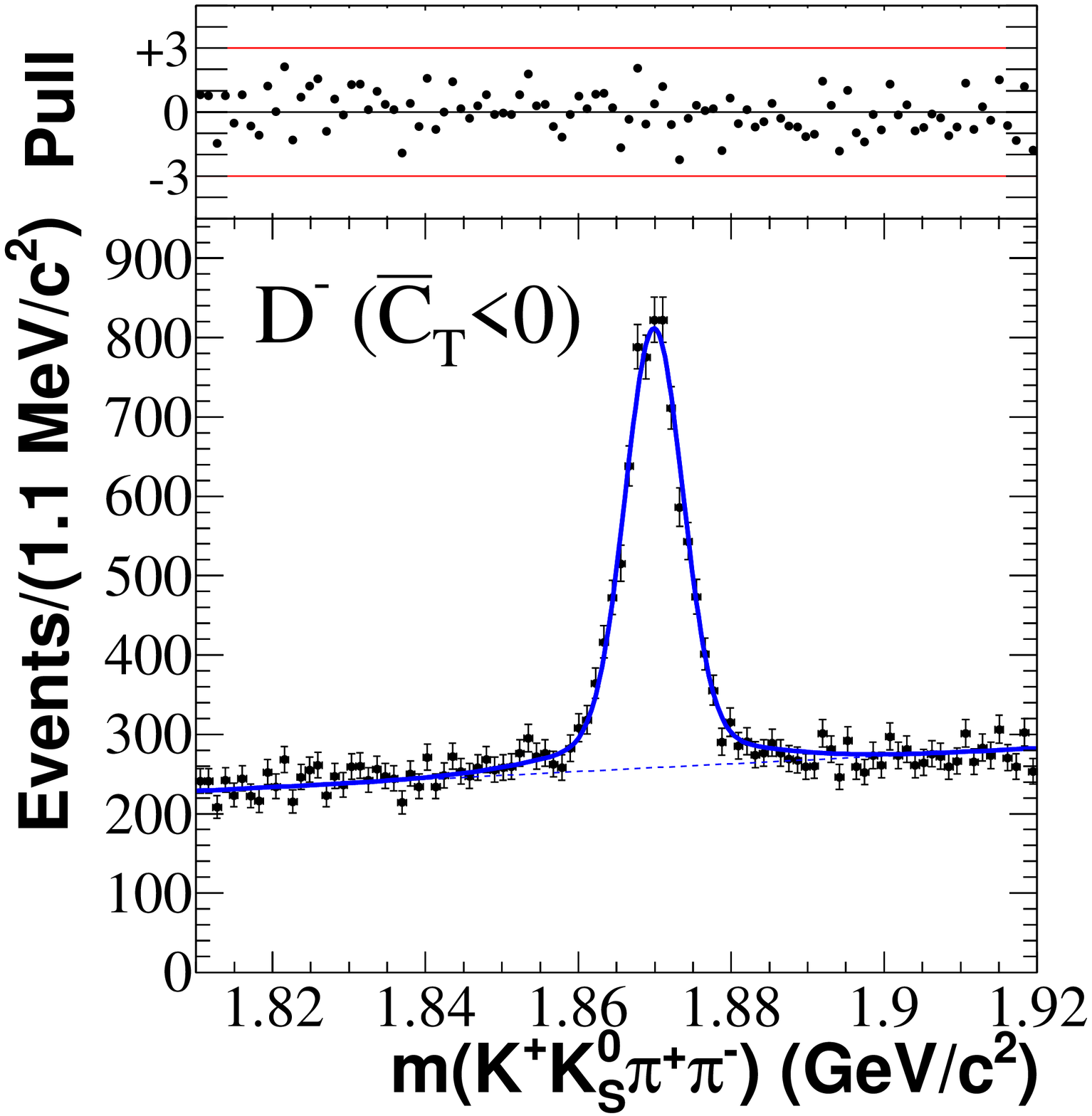}
\includegraphics[width=0.22\textwidth]{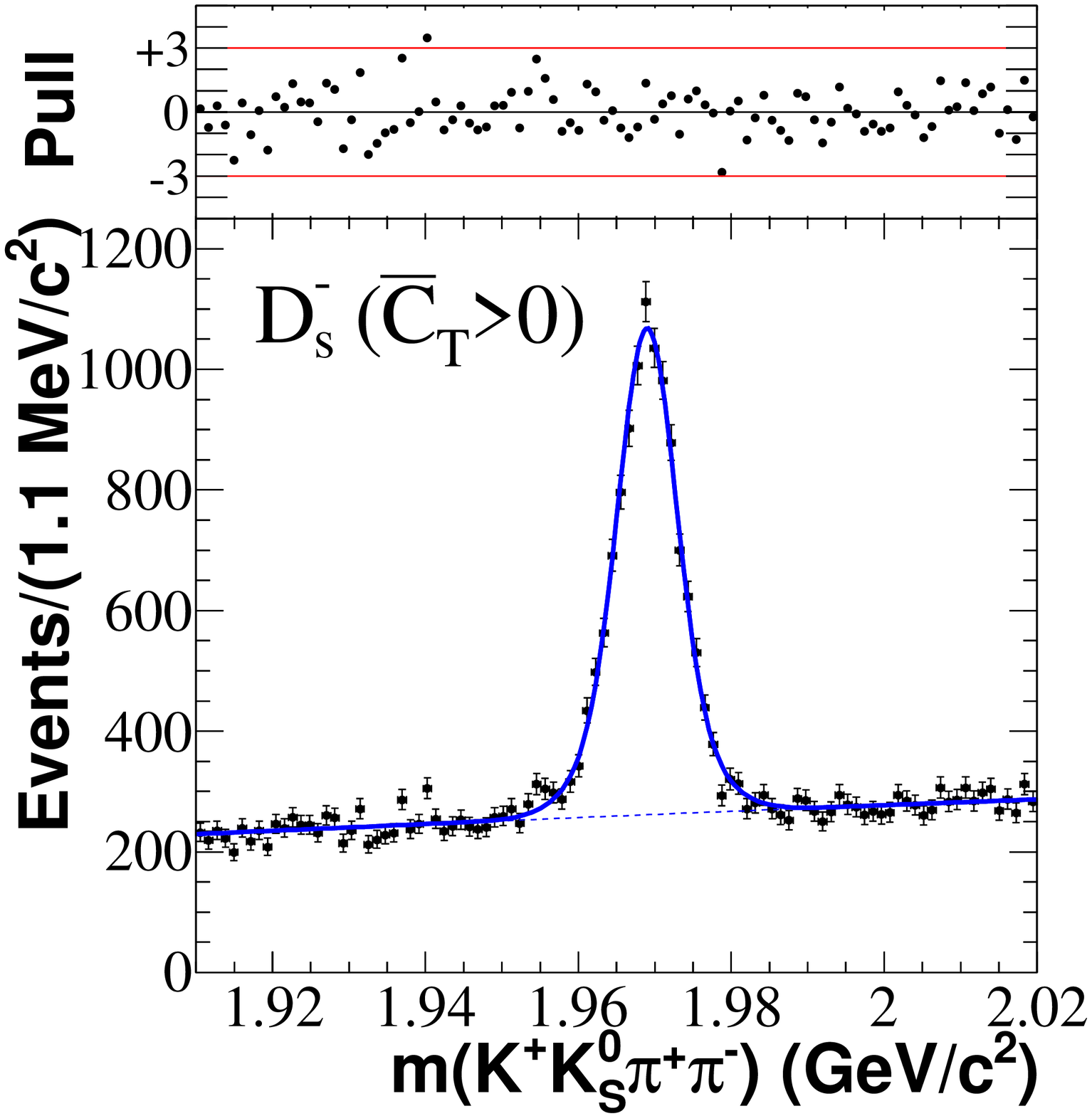}
\includegraphics[width=0.22\textwidth]{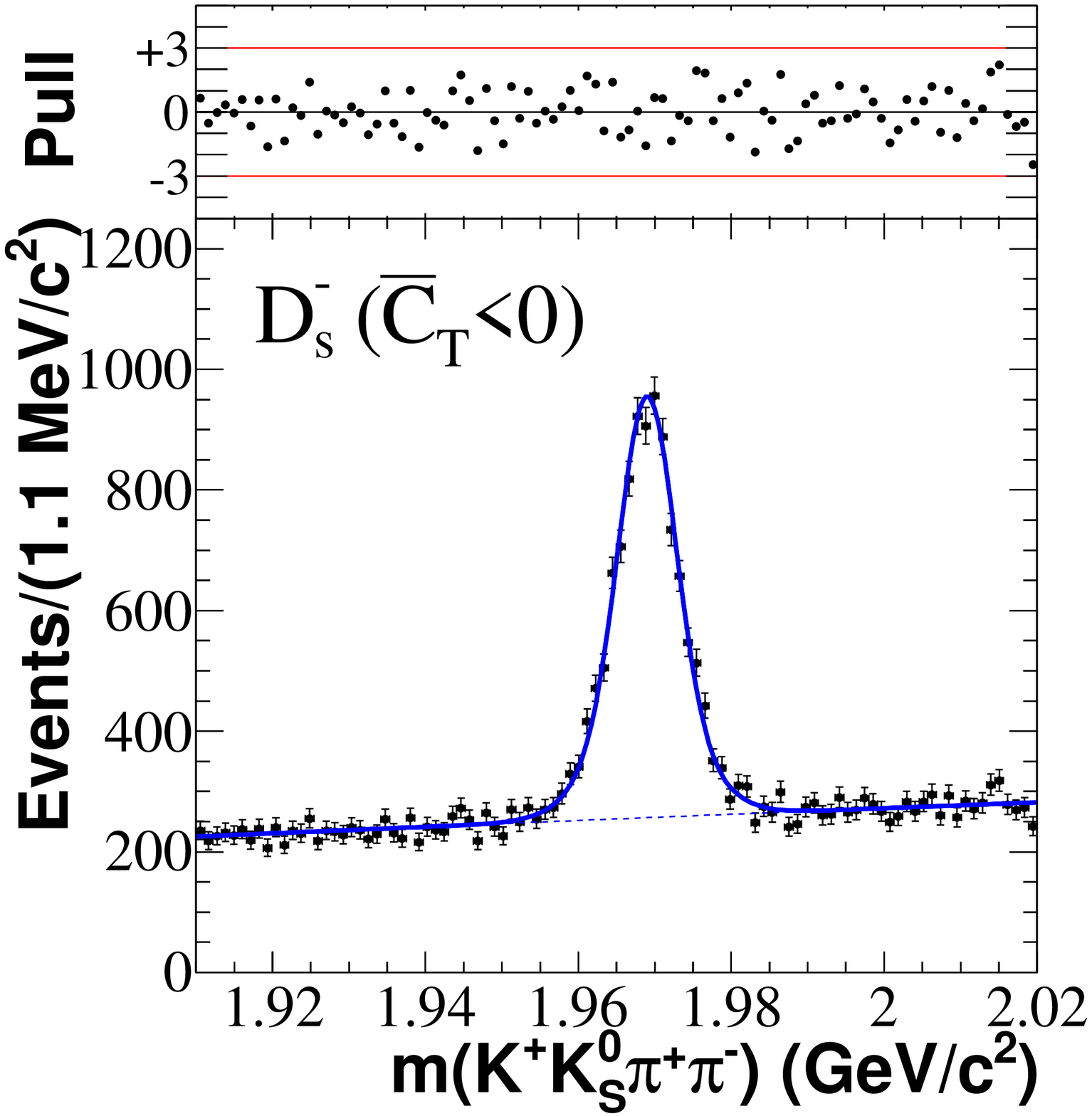}
\caption{\label{fig:Todd}The fit results for \Dp (left) and \Ds (right) decays, projected on the four subsamples separated using charge and $C_T$ ($\bar{C}_T$) value.}
\end{figure}

The results of the fits are shown in Fig.~\ref{fig:Todd} and give
\begin{align}
\nonumber 	A_T(\Dp)		&= (+11.2\pm14.1_{\text{stat}}\pm 5.7_{\text{syst}})\times10^{-3},\\
			\bar{A}_T(\Dm)	&= (+35.1\pm14.3_{\text{stat}}\pm 7.2_{\text{syst}})\times10^{-3},
\label{eq:AtRes}
\end{align}
and
\begin{align}
\nonumber 	A_T(\Ds)		&= (-99.2\pm10.7_{\text{stat}}\pm  8.3_{\text{syst}})\times10^{-3},\\
			\bar{A}_T(\Dsm) &= (-72.1\pm10.9_{\text{stat}}\pm 10.7_{\text{syst}})\times10^{-3}.
\label{eq:AtbRes}
\end{align}
Using Eq.~(\ref{eq:Atv}) we obtain the $T$ violation parameter values:
\begin{align}
\mathcal{A}_T(\Dp) &= ( -12.0 \pm  10.0_{\text{stat}} \pm 4.6_{\text{syst}} ) \times 10^{-3}
\end{align}
and
\begin{align}
\mathcal{A}_T(\Ds) &= ( -13.6 \pm 7.7_{\text{stat}} \pm 3.4_{\text{syst}} ) \times10^{-3}.
\end{align}

It can be noticed that the effect of FSI is larger for \Ds rather than \Dp decays. 
Such an effect is studied in detail in~\cite{Gronau:2011cf}.
However, the $T$ violation parameter is consistent to zero within the errors for both the two decay modes.

\section{Conclusion}
The search for CP violation in the charm sector has explored many channels and different approaches.
In the last years a vivid interest on this topic resulted into the publication of many new results.
We have reached the limit of the $B$ factories, obtaining sensitivities of $10^{-3}$, but the \CP violation from $c\to s$ transition did not show up yet, neither from SM or NP.
The new high-luminosity machines can light on this topic, providing new limits for \CP violation in charm decays or even a measurement.
We then strongly suggest to perform similar analysis at LHCb and to include them in the physics program of the next high-luminosity $B$ factories (SuperB and Super KEK-B).

\end{document}